# A Novel Hybrid Framework for Hourly PM$_{2.5}$ Concentration Forecasting Using CEEMDAN and Deep Temporal Convolutional Neural Network

**Fuxin Jiang[a, b], Chengyuan Zhang[c, *], Shaolong Sun[d, *], Jingyun Sun[e]**

[a]Academy of Mathematics and Systems Science, Chinese Academy of Sciences, Beijing 100190, China;
[b]School of Economics and Management, University of Chinese Academy of Sciences, Beijing 100190, China;
[c]School of Economics and Management, Beihang University, Beijing 100191, China;
[d]School of Management, Xi'an Jiaotong University, Xi'an, 710049, China;
[e]School of Statistics, Lanzhou University of Finance and Economics, Lanzhou 730020, China.

**Abstract**: For hourly PM$_{2.5}$ concentration prediction, accurately capturing the data patterns of external factors that affect PM$_{2.5}$ concentration changes, and constructing a forecasting model is one of efficient means to improve forecasting accuracy. In this study, a novel hybrid forecasting model based on complete ensemble empirical mode decomposition with adaptive noise (CEEMDAN) and deep temporal convolutional neural network (DeepTCN) is developed to predict PM$_{2.5}$ concentration, by modelling the data patterns of historical pollutant concentrations data, meteorological data, and discrete time variables' data. Taking PM$_{2.5}$ concentration of Beijing as the sample, experimental results showed that the forecasting accuracy of the proposed CEEMDAN-DeepTCN model is verified to be the highest when compared with the time series model, artificial neural network, and the popular deep learning models. The new model has improved the capability to model the PM$_{2.5}$-related factor data patterns, and can be used as a promising tool for forecasting PM$_{2.5}$ concentrations.

[*] Corresponding author. School of Economics and Management, Beihang University, Beijing 100191, China. Email: cyzhang@buaa.edu.cn (C. Y. Zhang).
[*] Corresponding author. School of Management, Xi'an Jiaotong University, Xi'an, 710049, China. Email: sunshaolong@xjtu.edu.cn (S. L. Sun).

# 1. Introduction

PM$_{2.5}$ (particles with an aerodynamic diameter of less than 2.5 μm) as the major pollutant in the atmosphere has attracted much attention in the field of forecasting, due to its negative impact on ambient air quality, public health, and socio-economic development (Miskell et al., 2019). Particularly, PM$_{2.5}$ concentration prediction is of great significance for controlling and reducing air pollution, which thus can help the government make effective early warning decisions and reminding the public to travel healthily (Bai et al., 2019; Shang et al., 2019). Therefore, more accurate prediction results of the PM$_{2.5}$ concentration prediction generated by an effective forecasting model will become more and more important (Gao et al., 2017).

It is worth noting that as a difficult task in the field of forecasting, the main reason for the difficulty of PM$_{2.5}$ forecasting can be attributed to the following two points. On one hand, the non-linear and non-stationary data characteristic of PM$_{2.5}$ time series affected by various external factors are the main obstacles for improving the prediction accuracy (Ma et al., 2020; Pak et al., 2020; Xu et al., 2020). Accordingly, the commonly related factors, including the meteorological factors (Jiang et al., 2017; Ni et al, 2017; Yang et al., 2018; Yang et al., 2020), geographical factors (Hu et al., 2013; Song et al., 2014) and time features (Zhang et al., 2014; Xu et al., 2020), are widely regarded as the external factor that make the corresponding time series present non-linear and non-stationary characteristics (Feng et al., 2015; Qi et al., 2019). On the other hand, PM$_{2.5}$ prediction involves high frequency of data, the wide range of sources of PM$_{2.5}$, multiple exogenous variables, which may cause problems such as long training time and large amount of data (Qi et al., 2019).

Under this background, a variety of data-driven methods have been developed to improve the prediction accuracy by dealing with the above mentioned obstacles (Combarro, 2013; Biancofiore et al., 2017; Wang et al., 2017; Shang et al., 2019). In general, depending on the way to construct the data-driven model, the previous prediction models are divided into two groups: linear forecasting and nonlinear forecasting models. As for linear forecasting model, geographically and temporally weighted regression model (Guo et al., 2020), multiple linear regression (MLR) (Moisan et al., 2018) and autoregressive integrated moving average models

(ARIMA) (Nieto et al., 2018) are widely introduced to forecast the concentration due to the interpretability of estimated parameters. However, the linear forecasting model cannot efficiently fit and model the nonlinearity and non-stationary of the $PM_{2.5}$. Moreover, these methods involving thousands of parameters requires prohibitive labor and computing resources to estimate the parameters (Qi et al., 2019).

Furthermore, the nonlinear forecasting models such as machine learning and deep learning are also popularly employed to model the data features of the $PM_{2.5}$, such as artificial neural networks (ANN) (Biancofiore et al., 2017), support vector regression (SVR) (Combarro, 2013; Nieto et al., 2013), extreme learning machine (Wang et al., 2017; Shang et al., 2019) and the long short-term memory neural network (LSTM) (Zhao et al., 2019; Wen et al., 2019). In particular, the deep learning neural networks (e.g., LSTM) are generally superior to the traditional machine learning methods in achieving excellent predictive performance (Zhao et al., 2019; Wen et al., 2019). The main reason is that deep learning methods can not only have the powerful modeling capability for more exogenous variables, but also effectively capture the long-term features and short-term features at the same time (Zhao et al., 2019). However, the commonly used deep learning such as Recurrent Neural Network (RNN) (Ong et al., 2016; Du et al., 2019), Deep Belief Network (Xing et al., 2020) and RNNs such as LSTMs may exists problems such as missing information based on "Forget Gate" due to network structure (Deng et al., 2019).

By contrast, the Temporal Convolutional Neural Network (TCN) as the generic architecture for convolutional sequence prediction can effectively capture all the information of the historical observations and the related discrete time variable, from the perspectives of the causal of the convolutions in the architecture (Chen al et., 2020). Notably, the decomposition method can effectively decompose the non-linear and non-stationary data into different timescales series, because various external factors affect $PM_{2.5}$ concentration on different time scales (Niu et al., 2017; Zhu et al., 2018; Liu et al., 2019). Therefore, this paper aims to introduce the Deep Temporal Convolutional Neural Network (DeepTCN) and complete ensemble empirical mode decomposition with adaptive noise (CEEMDAN) into the hourly $PM_{2.5}$ concentration modeling and forecasting. Particularly, the DeepTCN combined with the skip connection operation to make the network deeper can improve the prediction accuracy, by capturing the longer-term trend characteristics of $PM_{2.5}$ time series and the related factors time series such as weather.

The main contributions of the present study are as follows:

(1) a novel hybrid forecasting methodology with the decomposition method and deep learning method is proposed, in which the extracted multi-scale component of $PM_{2.5}$ concentrations is investigated;

(2) it is the first attempt to apply DeepTCN to $PM_{2.5}$ prediction, by capturing all $PM_{2.5}$-related factors' information and modelling the data patterns, in terms of the historical information of the pollutant concentrations, meteorological factors, and time variables;

(3) taking Beijing as a study sample, the proposed hybrid framework makes a better predictive performance than the other benchmarks involve time series method, artificial neural network method and deep learning methods, in terms of three popular evaluation criteria.

The remaining part of this paper is organized as follows. **Section 2** describes the proposed methodology. **Section 3** designs the empirical study. **Section 4** conducts the empirical results and discusses the effectiveness of the proposed method. **Section 5** concludes the paper and offers major directions for the future research.

## 2. Methodology

In this section, we propose a novel hybrid forecasting method, in which the CEEMDAN and deep learning technique (i.e., DeepTCN) are employed, for forecasting the $PM_{2.5}$ concentration, by modelling the different types of exogenous variables and capturing the corresponding complex data characteristics.

## 2.1 The framework

The general framework of the proposed method is described in **Fig. 1**. The framework describes the sub-steps including data collection, data processing, data decomposition, data forecasting, and results analysis.

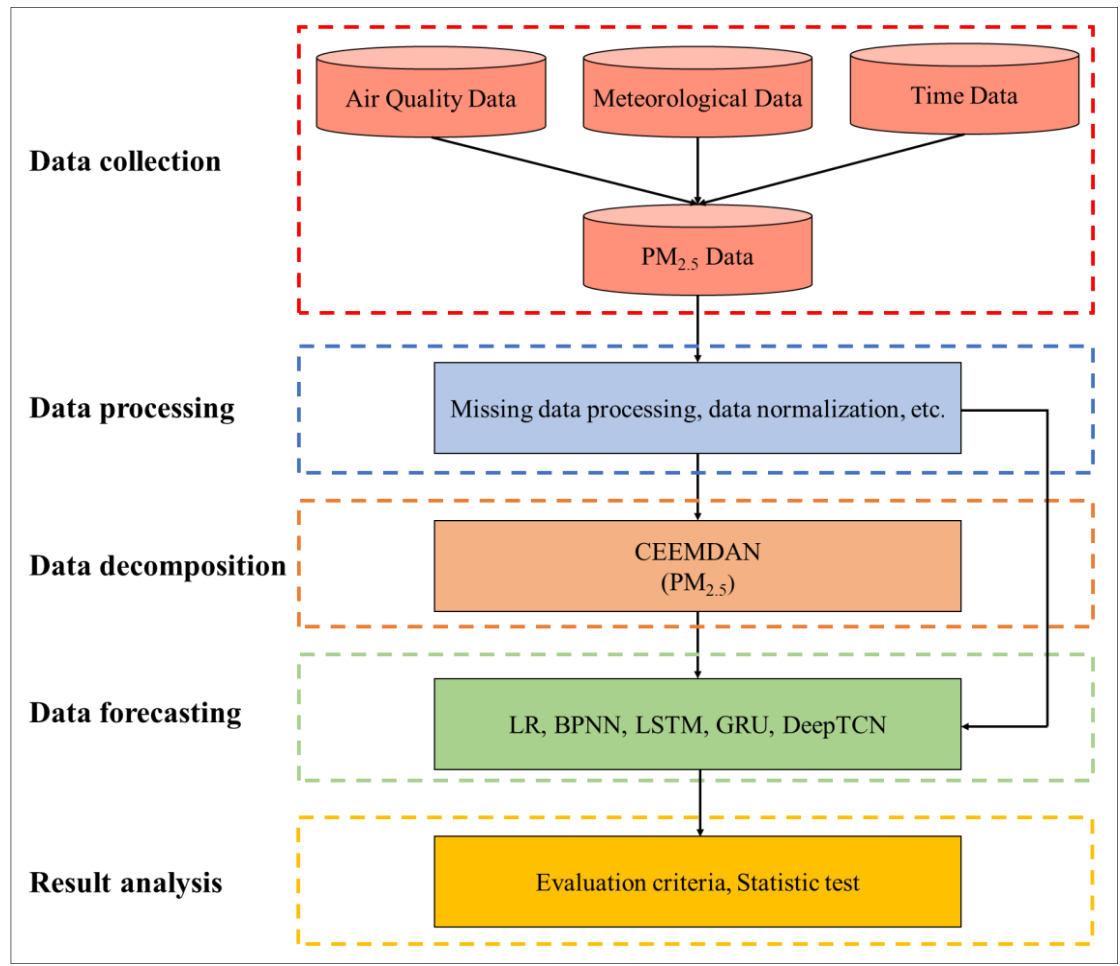

**Fig. 1** Proposed conceptual framework of $PM_{2.5}$ forecasting.

**Step 1: Variables collection**

In this step, the forecasting target data and corresponding variables' time series are collected with two sub-steps, i.e., variables selection and data collection. Accordingly, based on the previous literatures on $PM_{2.5}$ prediction, the mainly pollutant concentrations and the related meteorological factors are selected. Second, the different time series of the above continuous variables are respectively collected from the official website.

**Step 2: Data processing**

In this step, three sub-steps, i.e., missing data processing, data standardization and the categorical variables coding, are included for processing all the variables. First, due to data collection facility may malfunction, some data may be null. Therefore, the linear interpolation method is used to process the missing data to obtain a continuous time series. Second, all the continuous variables are standardized. Third, the categorical variables are encoded by label encoder for the deep neural network, for example, the sunny in weather variable is coded into the number 0, the fine with occasional clouds is coded into the number 1. The categorical variables are encoded by one-hot encoder for the linear regression model, for example, the sunny in weather variable is coded as (1,0,0,…,0), the fine with occasional clouds is coded as (0,1,0,…0).

**Step 3: Data decomposition**

In this step, the effective data decomposition technique is introduced for decomposing the $PM_{2.5}$ time series for extracting and modelling the non-linear and non-stationary data features, i.e., CEEMDAN. In particular, the $PM_{2.5}$ time series $x(t)$ is decomposed into $n-1$ intrinsic mode function (IMF) components $c_{j,t}$ ($j = 1, 2, ..., n-1$) and one residual ($r_t$). Correspondingly, each decomposition component (i.e., IMF and residual) includes specific timescale information, and the residual series reflects the trend of the original $PM_{2.5}$ data.

**Step 4: Data forecasting**

In order to explore the forecasting performance of the proposed CEEMDAN-DeepTCN method for $PM_{2.5}$, a comprehensive forecasting is designed in this step. Specifically, four popular forecasting techniques are used as the benchmarks to forecast the $PM_{2.5}$ (i.e., LR, BPNN, LSTM and GRU). Furthermore, all the three different type exogenous variables (i.e., pollutant concentrations, the meteorological factors, and time variables) were put into the model for forecasting $PM_{2.5}$.

**Step 5: Results analysis**

In this step, three sub-steps, i.e., results comparison, statistic test and robustness analysis, are included for analyzing the forecasting results of all the constructed forecast models. First, we calculate evaluation criteria (i.e., MAPE, RMSE, MAE) to compare forecasting performance of models and provide insights on PM$_{2.5}$ forecasting. Second, the statistic test, i.e., DM test, is employed for demonstrating the validity of the proposed method. Third, a robustness analysis is conducted for verifying the proposed methods' predictive stability.

**2.2 Complete ensemble empirical mode decomposition with adaptive noise**

The method of CEEMDAN (Torres et al., 2011) is an expansion algorithm of ensemble empirical mode decomposition (EEMD) by adding adaptive Gaussian white noise to smooth pulse interference in data decomposition, by the following process:

1) A collection of Gaussian white noise series is added to the original data as:

$$s^i(t) = x_t + \varepsilon_i v^i(t), \qquad (1)$$

where $s^i(t)$ is the time series with the additional noise in the $i$th trial ($i=1,2,\ldots,I$), $x_t$ is the original data, $\varepsilon_i$ is the ratio of the additional noise to the original signal, and $v^i(t)$ is the Gaussian white noise series.

2) EMD is used to obtain the first intrinsic mode function $I_1(t)$ as:

$$I_1(t) = (\sum_{i=1}^{I} I_1^i(t))/I, \qquad (2)$$

where $I_1^i(t)$ is the first intrinsic mode function obtained in the $i$th trial.

3) The first residue $r_1(t)$ is as:

$$r_1(t) = x_t - I_1(t), \qquad (3)$$

4) The $k$th residue is as:

$$r_k(t) = r_{k-1}(t) - I_k(t), \qquad (4)$$

where $I_k(t)$ is the $k$th intrinsic mode function of CEEMDAN.

5) The $I_{k+1}(t)$ of CEEMDAN is:

$$I_{k+1}(t) = \sum_{i=1}^{I} E_1(r_k(t) + \varepsilon_k E_k(v^i(t)))/I, \qquad (5)$$

where $E_k(\cdot)$ is the *k*th mode of EMD.

Repeat steps 4) and 5) until all the intrinsic mode functions are found.

**2.3 Deep TCN Model**

Deep Temporal convolution network (DeepTCN) is a deep learning technique combined with one-dimensional convolution that can be used to model sequence data, and the empirical study on multiple datasets shows the framework outperforms generic recurrent architectures such as LSTMs and GRUs. Based on the above, we propose a deep temporal convolution network (DeepTCN) model with Embedding as shown in **Fig. 2**.

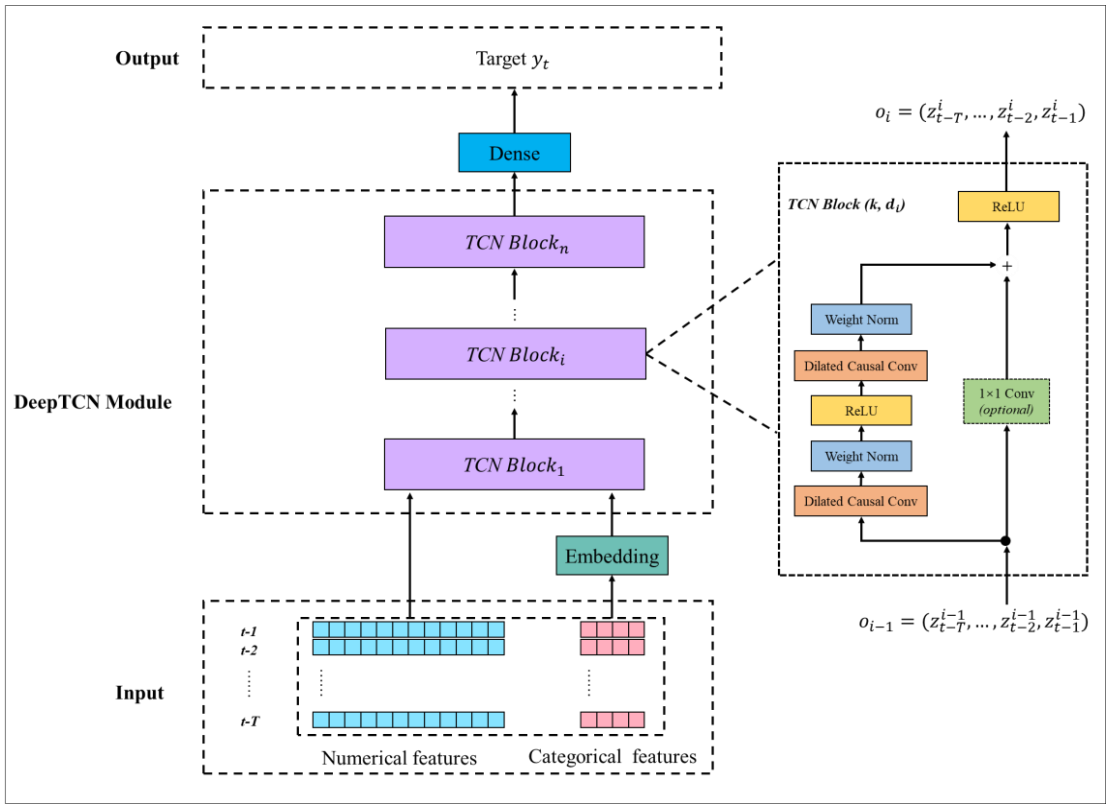

**Fig. 2** Architecture of DeepTCN.

***Input.*** The input consists of two parts, one is numerical features $X_{num} = \{x_t^{num}\}_{t=1}^{N}$ ($x_t^{num} \in R^{num}$), *num* represents continuous variables, including $PM_{2.5}$, $PM_{10}$, $NO_2$, $SO_2$, $O_3$, CO, horizontal wind speed, vertical wind speed, temperature, precipitation, pressure, relative humidity, and IMFs and residue decomposed by CEEMDAN, and the other is categorical features $X_{cat} = \{x_t^{cat}\}_{t=1}^{N}$ ($x_t^{cat} \in R^{cat}$), *cat* represents discrete variables, including month of year, day of week, hour of day and weather, where *N* is the number of samples.

*Embedding.* In order to unify the numerical features and categorical features, we use embedding operation to convert categorical features into vectors of dimension. This operation is similar to word embedding in natural language processing tasks, which can be trained with the network. In this way, the categorical features have a "semantic" meaning and can be directly input into the neural network.

$$\hat{X}_{cat} = Embedding(X_{cat}), \tag{6}$$

Then, we concatenate the vector of categorical features after embedding with numerical features as the input for the next stage.

$$\hat{X} = Concat(X_{num}, \hat{X}_{cat}), \tag{7}$$

*DeepTCN Module.* As shown in **Fig. 2**, the DeepTCN module is formed by stacking multiple TCN blocks, and each TCN block contains two causal dilated convolution layers, where *k* is the size of kernel and $d_i$ is the dilation factor in the *i*-th TCN Block. Causal convolutions are convolutions where an output at time $t'$ can be only obtained from inputs that are no later than $t'$ and using dilation strategy can capture much longer sequences. Given the input sequence $\hat{X} = \{\hat{x}_t\}_{t=1}^{N}$ $\hat{x}_t \in R^{num+cat}$, the output *s* at location $t'$ of a causal dilated convolution with kernel $\omega$ and can be expressed as:

$$s_{t'} = (\hat{X} *_d \omega)(t') = \sum_{l=0}^{k-1} \omega(l) \cdot \hat{x}_{t'-d \cdot l}, \tag{8}$$

where *d* is the dilation factor, and *k* is the size of the kernel. We concatenate *K* kernel functions' outputs together to get more information as follow:

$$x_{t'} = [s_{t'}^1, s_{t'}^2, ..., s_{t'}^K], \tag{9}$$

where *K* is the number of kernel functions. Stacking multiple dilated convolutions enable networks to have very large receptive fields and to capture long-range temporal dependencies with a smaller number of layers.

In order to enhance the robustness of model training, we add a weight normalization layer (Salimans et al., 2016) after causal dilated convolution layer. Weight normalization is to reparametrize the original weight matrix $W$ of causal dilated convolution as follow:

$$W = \frac{g}{\|V\|} V, \tag{10}$$

where $V$ is matrix that is the same as the shape of $W$, $g$ is a scalar. Backpropagation is used to update $g$ and $V$.

Then a rectified linear unit (ReLU) activation is applied followed by weight normalization layer, the rectified linear unit activation helps to avoid gradient vanishing in the deep networks. The rectified linear unit activation as follows:

$$ReLU(x) = max(0, x), \tag{11}$$

Another weight normalization layer is added after the second causal dilated convolution layer. The output after the second weight normalization is added to the input of the TCN block and the addition is then followed by a second ReLU activation as follow:

$$o = ReLU(x + \mathcal{F}(x)), \tag{12}$$

where $\mathcal{F}$ is the above series of transformations consisting of causal dilated convolution, weight normalization and ReLU.

In order to make the output and input shape the same, we add $1 \times 1\ convolution$ to the input. Eq. (12) can be rewritten as:

$$o = ReLU(conv_{1 \times 1}(x) + \mathcal{F}(x)), \tag{13}$$

**Dense.** For the output of the DeepTCN module stacked by $n$ TCN blocks $o_n = (z_{t-T}^n, ..., z_{t-2}^n, z_{t-1}^n)$, we use a fully connected layer to transform the last time step $z_{t-1}^n$ to get the final result as follow:

$$\hat{y}_t = W_y z_{t-1}^n + b_y, \tag{14}$$

where $W_y$ and $b_y$ are the weights and bias of the fully connected layer.

## 2.4 Benchmarks

In this paper, we compare the proposed forecasting model with several benchmarks including the time series model, machine learning model and the common deep learning models.

**(1) LR**

LR is a most basic time series method for forecasting, as follows:

$$y = \beta_0 + \beta_1 x_1 + \beta_2 x_2 + \cdots + \beta_k x_k + \mu, \tag{15}$$

where $x_i$, ($i=1,\ldots,k$) is the $i$th input data, $\beta_j, (j=0,1,\ldots,k)$ are the trained coefficients, $y$ is the prediction dependent variables, and $\mu$ is the errors.

**(2) BPNN**

BPNN, a typical case of neural networks, has widely been employed in the field of forecasting (Sun et al., 2019). BPNN takes two steps, i.e., the positive propagation of information and the reverse propagation of error, to recursively adjust the model parameters and mitigate the modeling error to an acceptable level (Rumelhart and McClelland, 1986). A basic BPNN includes an input layer, a hidden layer, and an output layer, with their relationships mathematically described as (Rumelhart and McClelland, 1986):

$$h_j = f_1(\sum_{i=1}^{n} w_{j,i} x_i + \theta_j), (\theta_j \geq 0, w_{j,i} \leq 1), \tag{16}$$

$$y = f_0(\sum_{j=1}^{m} w_{0,j} h_j + \lambda_0), (\lambda_0 \geq 0, w_{0,j} \leq 1), \tag{17}$$

where $x_i$ is the $i$th input layer node, meanwhile, in Eq. (16) and Eq. (17), $h_j$ and $y$ represent the, the $j$th hidden layer node and the output respectively. $\theta_j$ and $\lambda_0$ are the biases in the hidden and output layers, respectively. $n$ and $m$ are the total numbers of the nodes in the input and hidden layers, respectively. $w_{j,i}$ and $w_{0,j}$ are the weights for the hidden and output layers. As for the activation functions, $f_1(\cdot)$ and $f_0(\cdot)$ are respectively in the hidden and output layers. Here, the model parameters, i.e., the bias and weights, are recursively trained via an iteratively searching method (e.g., gradient descent).

**(3) LSTM**

Unlike a single neural network layer, LSTM is composed of an input gate $i_t$, a forget gate $f_t$, an output gate $o_t$ and a cell $C_t$. The design of the gates in LSTM partially avoids exploding and vanishing gradient problems. The architecture of a set of LSTM units is as follows (Li et al., 2017):

$$f_t = \sigma(W_f \cdot [h_{t-1}, x_t] + b_f), \tag{18}$$

$$i_t = \sigma(W_i \cdot [h_{t-1}, x_t] + b_i), \tag{19}$$

$$\tilde{C}_t = \tanh(W_C \cdot [h_{t-1}, x_t] + b_C), \tag{20}$$

$$C_t = C_{t-1} \odot f_t + i_t \odot \tilde{C}_t, \tag{21}$$

$$o_t = \sigma(W_o \cdot [h_{t-1}, x_t] + b_o), \tag{22}$$

$$h_t = o_t \odot \tanh(C_t), \tag{23}$$

$$\hat{y}_t = \sigma(W_{fc} h_t + b_{fc}), \tag{24}$$

where $W_f$, $W_i$, $W_o$, $W_C$ and $W_{fc}$ denote weights matrices of the input-hidden layer, hidden-output layer; $b_f$, $b_i$, $b_o$, $b_C$ and $b_{fc}$ denote vectors of biases; and $\sigma$ denotes the sigmoid activation function.

**(4) GRU**

GRU is a widely-used and simpler variant of the LSTM unit. It simplifies the structure with two gates: the update gate $z_t$ and the reset gate $r_t$. The following formula is derived from the GRU model (Du et al., 2019; Shahid et al., 2020):

$$r_t = \sigma(W_r \cdot [h_{t-1}, x_t]), \tag{25}$$

$$z_t = \sigma(W_z \cdot [h_{t-1}, x_t]), \tag{26}$$

$$\tilde{h}_t = \tanh(W_{\tilde{h}} \cdot [r_t * h_{t-1}, x_t]), \tag{27}$$

$$h_t = (1 - z_t) * h_{t-1} + z_t * \tilde{h}_t, \tag{28}$$

$$\hat{y}_t = \sigma(W_{fc} h_t + b_{fc}), \tag{29}$$

where $z_t$ is the update gate; $r_t$ is the reset gate; the activation $h_t$ is a linear interpolation between the candidate activation $\tilde{h}_t$ and the previous activation $h_{t-1}$.

## 3. Empirical design

For illustration and verification, the proposed CEEMDAN-DeepTCN methods is used to predict the PM$_{2.5}$ concentration of a highly air-polluted city in Beijing. Correspondingly, the pollutant concentrations (i.e., PM$_{2.5}$, PM$_{10}$, NO$_2$, SO$_2$, O$_3$, CO), the meteorological factors (wind speed, wind direction, temperature, precipitation, pressure, relative humidity, weather) and temporal information are given to validate the wide applicability of the hybrid model and ensure the objectivity of the experiments. The following subsections describe the study data, set the evaluation criteria and display the corresponding forecasting technique parameters, respectively.

### 3.1 Data descriptions

To validate the effectiveness of the proposed forecasting framework, the data investigated and collected in this study are hourly historical observations of PM$_{2.5}$ concentration, meteorological variables and the time variables in Beijing, China. As shown in **Fig. 3**, PM$_{2.5}$ concentrations is collected from environmental monitoring station that located at 116º24´E and 39º54', and **Table 1** summarizes the statistics of related pollutant variables. The hourly dataset of pollutants (PM$_{2.5}$, PM$_{10}$, NO$_2$, SO$_2$, O$_3$, CO) during 2 January 2015 to 31 December 2017 (totally 26,280 h) are collected from the website https://quotsoft.net/air/. Moreover, the horizons of prediction are set to 1-3 hour(s) to verify the robustness of the proposed forecasting model.

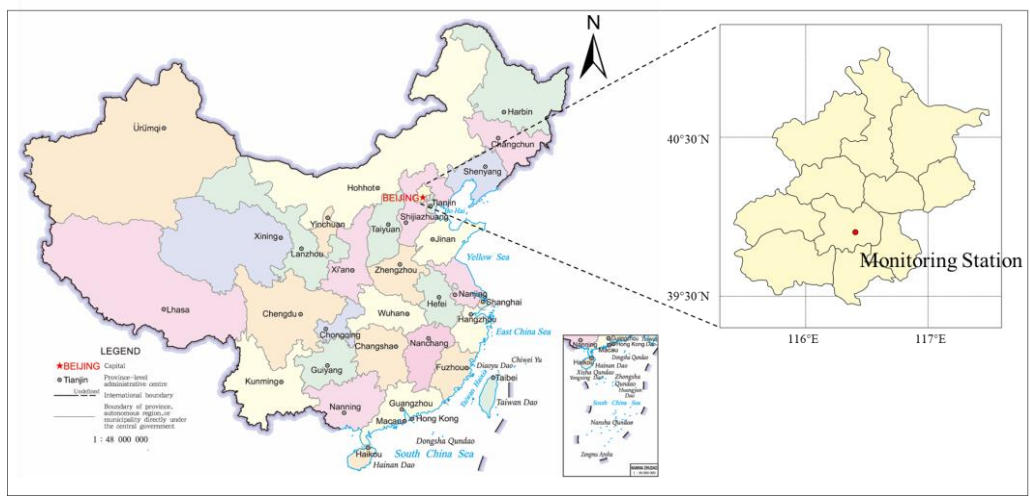

**Fig. 3** Geographic locations of Beijing's monitoring station.

**Table 1.** Statistics of the pollutant and meteorological variables.

| Variable | Unit | Range | Mean | St.dev. |
|---|---|---|---|---|
| $PM_{2.5}$ | μg/m³ | [2,692] | 69 | 71.81 |
| $PM_{10}$ | μg/m³ | [1,1000] | 97.82 | 88.33 |
| $NO_2$ | μg/m³ | [2,192] | 45.6 | 27.69 |
| $SO_2$ | μg/m³ | [1,248] | 9.72 | 12.65 |
| $O_3$ | μg/m³ | [1,339] | 59.03 | 54.39 |
| CO | μg/m³ | [0.13, 9.63] | 1.12 | 1.06 |
| Wind_x | km/h | [-44, 48.86] | 1.17 | 11.45 |
| Wind_y | km/h | [-54.16, 44.43] | -0.39 | 8.65 |
| Temperature | °C | [-17,46] | 14.83 | 12.24 |
| Precipitation | mm | [0,251.7] | 0.26 | 4.06 |
| Pressure | hPa | [992,1047] | 1017.16 | 10.65 |
| Relative humidity | % | [5,97] | 42.69 | 22.43 |

Moreover, six types of hourly meteorological data, i.e., wind speed, wind direction, temperature, precipitation, pressure, relative humidity are collected in this study. Detailed information of the related meteorological variables is shown in **Table 1**. The reasons for choosing the above meteorological factors mainly come down to two points. First, the temperature and the pressure have effect on the atmospheric and ventilation conditions, the humidity, and the precipitation have impact on the deposition characteristic of particle matter, and the wind speed assists in the dispersion of particle matter (Bai et al., 2019). Second, the previous literature has achieved good prediction performance by introducing the above meteorological data into the prediction model, which proves the effectiveness of the relevant meteorological indicators as forecasting factors (Jiang et al., 2017; Ni et al, 2017; Yang et al., 2018; Yang et al., 2020). In addition, for the weather variables, different weather patterns are demonstrated in **Fig. 4**. Accordingly, all the meteorological data (i.e., wind speed, temperature, precipitation, pressure, relative humidity, weather) are downloaded from the website https://www.heweather.com in the corresponding periods for this research.

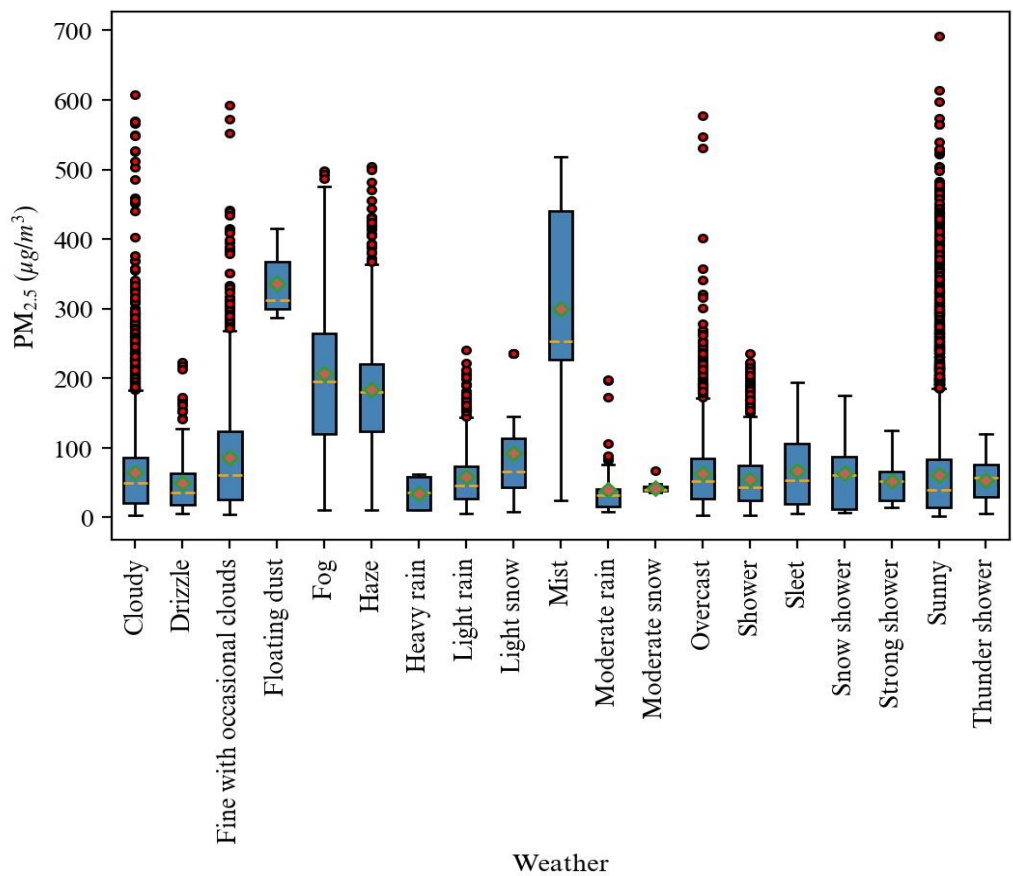

**Fig. 4** The distributions of different weather variables.

As for different time variables, the hourly, daily, and monthly variables are processed as the categorical features respectively, and then incorporated into incorporated into the prediction model as the exogenous variables. Furthermore, the corresponding violin plots reveals the distribution of different time variables at different frequencies from 2 January 2015 to 31 December 2017, as shown in **Fig. 5**. Specifically, as for the hourly variables, it can be seen that $PM_{2.5}$ in the early morning and evening is slightly smaller than that at noon and late at night. Among them, the reason for the high concentration at night and noon may be that the temperature prevents the effective diffusion of $PM_{2.5}$. This also means that different times of the day have different predictive capabilities. As for the daily variables, the $PM_{2.5}$ value on Wednesday, Friday, and Sunday is obviously larger than the other days. In addition, in the winter of November, December and January, the $PM_{2.5}$ in Beijing is significantly greater than that in the months of spring and summer, and show a "U" distribution as whole.

Therefore, these above exogenous variable patterns imply that the information of the time variables might at different frequencies have a great predictive power for the $PM_{2.5}$. In the past, the previous literature incorporated weather and time variables into the prediction model as much as possible at the same time to improve prediction accuracy (Maciąg et al., 2019; Xu et al., 2020). Hence, to effectively capture the pattern of the variables described above on $PM_{2.5}$ concentration prediction, we employ the embedding technology map the categorical variables of time and weather into vectors, and add them to the DeepTCN model for end-to-end training. **Table 2** summarizes the statistics of related categorical variables.

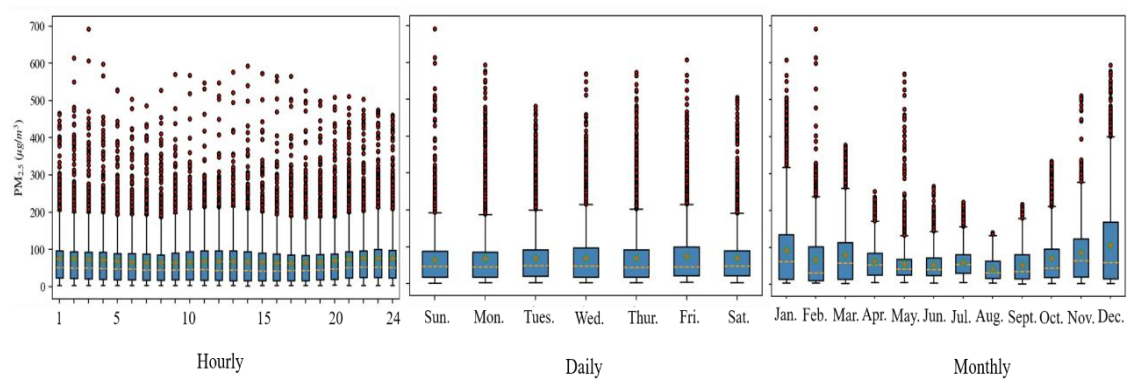

**Fig. 5** The distributions of different time variables at different frequencies.

**Table 2.** Statistics of categorical features.

| Variable | Category |
|---|---|
| Weather | Sunny, Fine with occasional clouds, Cloudy, Overcast, Light snow, Moderate snow, Snow shower, Sleet, Light rain, Drizzle, Shower, Strong shower, Thunder shower, Moderate rain, Heavy rain, Mist, Haze, Fog, Floating dust, Sand blowing |
| Monthly | January, February, March, April, May, June, July, August, September, October, November, December |
| Daily | Sunday, Monday, Tuesday, Wednesday, Thursday, Friday, Saturday |
| Hourly | 1,2,…,24 |

## 3.2 Evaluation criteria

To measure forecasting performance, the popular three criteria were adopted to evaluated the forecasting accuracy:

$$MAPE = 100\% \times \sum_{t=1}^{N} |1 - \hat{y}_t / y_t|, \qquad (30)$$

$$RMSE = \sqrt{\frac{1}{N} \sum_{t=1}^{N} (y_t - \hat{y}_t)^2}, \qquad (31)$$

$$MAE = \frac{1}{N} \sum_{t=1}^{N} |y_t - \hat{y}_t|, \qquad (32)$$

where $N$ is the size of the testing size $y_t$ and $\hat{y}_t$ are actual and predicted values at period t, respectively.

To verify the superiority of the proposed approach from a statistical perspective, the popular statistic method (i.e., Diebold-Mariano (DM) test) was adopted to test the statistical significance of all models using the mean absolute percent error as the loss function, with the null hypothesis that the comparison forecasting models appear a similar forecast accuracy (Diebold & Mariano, 2002). In this study, the DM statistic can be defined as:

$$DM = \frac{\bar{D}}{(\hat{V}_{\bar{D}} / N)^{1/2}}, \qquad (33)$$

where $\bar{D} = 1/N \sum_{t=1}^{N} \left( |1 - \hat{y}_{A,t} / y_t| - |1 - \hat{y}_{B,t} / y_t| \right)$, $\hat{V}_{\bar{D}} = \gamma_0 + 2 \sum_{k=1}^{h-1} \gamma_k$ ($\gamma_k = 1/N \sum_{t=|k|+1}^{N} (D_t - \bar{D})(D_{t-|k|} - \bar{D})$, $D_t = |1 - \hat{y}_{A,t} / y_t| - |1 - \hat{y}_{B,t} / y_t|$), $h$ is the forecast horizon and $\hat{y}_{A,t}$ and $\hat{y}_{B,t}$ denote the forecasts for $y_t$ generated by the proposed approach A and benchmark model B, respectively, at time $t$.

## 3.3 Model specification

Details of experimental settings are summarized in **Table 3**. The data set was spilt into 60% for train set (covering the period from 2 January 2015 to 31 October 2016, 16056 observations), 20% for validation set (covering the period from 1 November 2016 to 31 May 2017, with 5088 values) and 20% for test set (from 1 June 2017 to 31 December 2017, with 5136 values) to ensure the reliability of our proposed method. In BPNN, a usually three-layer network was built with 32 hidden nodes. In LSTM and GRU, the hidden size was set to 64. Our proposed framework employs the DeepTCN to capture the temporal correlations and embedding layer to map the categorical variables. The network is stacked by TCN blocks. In our experiment, the number of TCN blocks was set to 4, the dilation factors of four TCN blocks were set to {1, 2, 4, 8}, the number of kernel functions of four TCN blocks were set to {32, 32, 16, 16}, the kernel size of four TCN blocks was set to 2. Furthermore, four categorical variables were embedded into vector, the embedding size of month was set to 2, the embedding size of day was also set to 2, the embedding size of hour was set to 4 and the embedding size of weather was set to 2. MAPE has advantages of scale-independency and interpretability, MAPE was adopted as loss function. Adam was optimizer of our models. The training epochs was 100, the batch size was 128 and the learning rate was 0.01.

**Table 3.** Details of the experimental settings.

| Parameter | Value |
| --- | --- |
| Number of records | 26, 280 |
| Time interval (h) | 1 |
| Training set | 60% |
| Validation set | 20% |
| Test set | 20% |
| History length (T) | 24 |
| Training Epochs | 100 |
| Learning rate | 0.01 |
| Batch size | 128 |
| Optimizer | Adam |
| Loss function | Mean absolute percent error |
| BPNN hidden size | 32 |
| LSTM hidden size | 64 |
| GRU hidden size | 64 |
| Embedding size of Month | 2 |
| Embedding size of Day | 2 |
| Embedding size of Hour | 4 |
| Embedding size of Weather | 2 |
| Number of TCN blocks | 4 |
| Dilation factor of TCN blocks | {1,2,4,8} |
| Number of kernel functions of TCN blocks | {32,32,16,16} |
| Kernel size of TCN blocks | 2 |

## 4. Empirical results

For illustration and verification, the data decomposition results by the CEEMDAN-based method and the comparison of finally forecasting results for proposed method and benchmarks were presented in **Section 4.1**, **Section 4.2** and **Section 4.3**, respectively. In addition, a robustness analysis was conducted in **Section 4.4**, and **Section 4.5** summarized the major conclusions of the empirical results.

### 4.1 Decomposition results

The decomposition result of hourly $PM_{2.5}$ data of Beijing by using the CEEMDAN-based method was demonstrated for analyzing the effect of complex impact factors on $PM_{2.5}$. **Fig. 6** shows the IMF components and the residue, which were all extracted from the original data, were sorted from the highest frequency to the lowest frequency, respectively. In particular, fifteen IMFs and one residue were extracted from the original $PM_{2.5}$ concentration time series by using the CEEMDAN.

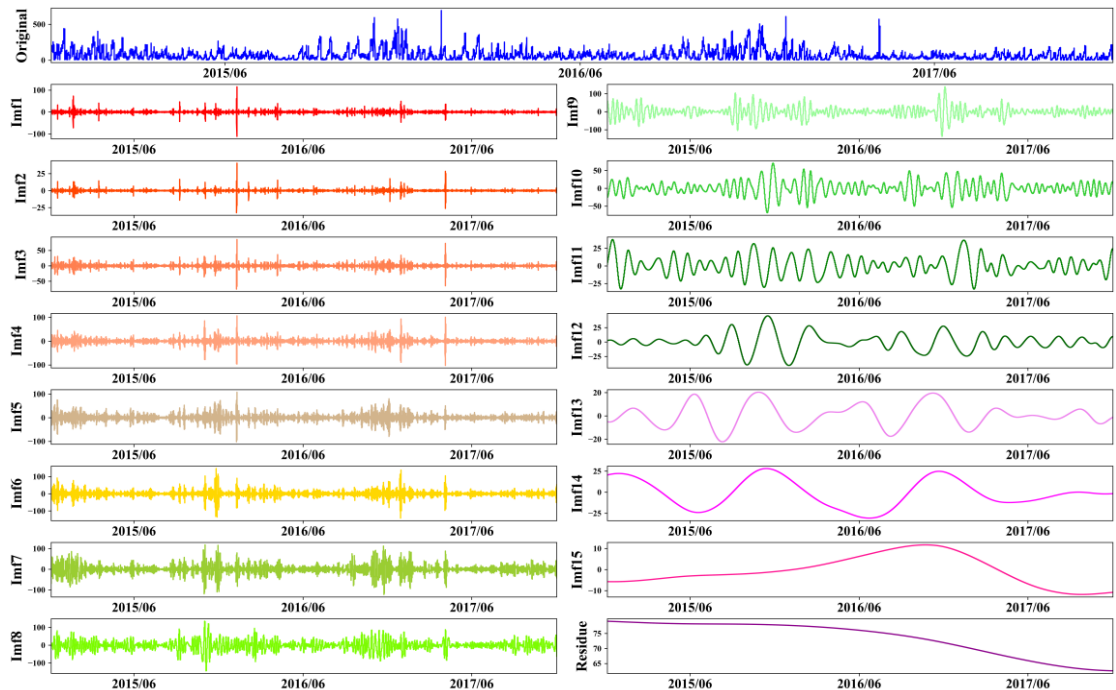

**Fig. 6** The decomposed components of $PM_{2.5}$.

## 4.2 Prediction results

Focusing on the forecasting results, a comprehensive comparison with four types of benchmarks is conducted. The **Table 4** shows the comparison result with one-, two-, and three-step-ahead forecasting from the perspective of MAPE, MAE and RMSE respectively, and **Table 5** demonstrates the DM test's result for verifying the validity of the proposed approach statistically. The bold font in **Table 4** shows the model with the lowest evaluation criteria at each horizon. From the evaluation criteria and the statistical test, one important conclusion can be clearly deduced that the proposed approach CEEMDAN-DeepTCN can be consistently proved to outperform the corresponding other forecasting models, at the confidence level of 95% base DM test.

Firstly, focusing on the MAPE in **Table 4**, we can see that the traditional time series model, i.e., LR, is the worst performer at all horizons with an overall average 28% of MAPE at forecasting $PM_{2.5}$ concentration, and as can be seen that this benchmark model made the worst forecasts in comparison with other advanced models. Interestingly, the deep learning techniques in this study (i.e., LSTM, GRU and the DeepTCN) are overall achieved a better prediction performance than the other two models at all horizons. Furthermore, the proposed CEEMDAN-DeepTCN prediction model made a most accurate forecasts among all the other five forecasting models with a low value on average across the horizons (4.916% MAPE). In particular, the MAPE of the CEEMDAN-DeepTCN model are approximately 83.01%, 80.62%, 72.93%, 72.71% and 70.37%, lower than those benchmarks using LR, BPNN, LSTM, GRU and DeepTCN, respectively, and thereby portrays CEEMDAN-DeepTCN's capabilities of providing comparatively stable and more accurate forecasts in the short run for forecasting $PM_{2.5}$ concentration.

It is evident from **Table 4** that based on the MAE criterion, CEEMDAN-DeepTCN outperforms LR, BPNN, LSTM, GRU and DeepTCN comfortably by recording the lowest forecasting error for forecasting $PM_{2.5}$ concentration at all horizons. Specifically, the MAE of the CEEMDAN-DeepTCN model are approximately 81.02%, 80.78%, 77.38%, 77.32% and 76.38% lower than those benchmarks using LR, BPNN, LSTM, GRU and DeepTCN at $h = 1$,

2 and 3 hours ahead respectively. Likewise, DeepTCN's performance is inferior only to CEEMDAN-DeepTCN, but it performs better than the other four methods, and belongs to the second place in predicted performance. In addition, the prediction accuracy of LSTM is the worst among the three methods of deep learning, which belongs to the third place in the prediction performance. As for the time series model and artificial neural network model, the prediction performance of these two methods is almost the same. When horizon=1 and 3, LR is better than BPNN, and when horizon = 2, it is just the opposite.

As for RMSE in **Table 4**, overall, based on the highest number of bold outcomes reported by a particular model we can suggest that on average across all horizons the employed deep learning forecasting model in this study, i.e., CEEMDAN-DeepTCN, is able to provide the optimal accuracy forecasts in comparison to forecasts from the other models. More specifically, the RMSE of the CEEMDAN-DeepTCN model are approximately 81.26%, 81.36%, 79.63%, 79.74% and 79.01% lower than those benchmarks using LR, BPNN, LSTM, GRU and DeepTCN, respectively. The comparison results of RMSE from **Table 4** show that on average, CEEMDAN-DeepTCN provides a better forecasting accuracy in comparison to the other five forecasting models for $PM_{2.5}$.

Taking a closer look at the three evaluation criteria (i.e., MAPE, MAE and RMSE in **Table 4**) for the monitoring station at each horizon in detail uncovers the following. Firstly, we find that the deep learning models provide the better forecasts for $PM_{2.5}$ concentrations at $h = 1, 2$ and 3 steps-ahead. Secondly, the proposed CEEMDAN-DeepTCN model outperforms all the corresponding benchmarks in all horizons, testifying to its effectiveness for $PM_{2.5}$ forecasting in terms of level forecasting accuracy. These results repeatedly support the effectiveness of the proposed deep learning model (i.e., CEEMDAN-DeepTCN) in modelling the data patterns and improving prediction accuracy, especially for $PM_{2.5}$ concentrations.

Table 4. Performance comparison of different methods in terms of criteria.

| $h$ | Criteria | Forecasting models | | | | | |
|---|---|---|---|---|---|---|---|
| | | LR | BPNN | LSTM | GRU | DeepTCN | CEEMDAN-DeepTCN |
| One | MAPE | 0.1402 | 0.1343 | 0.0965 | 0.0963 | 0.0920 | **0.0265** |
| | MAE | 3.4900 | 3.5870 | 3.0390 | 2.9800 | 2.8290 | **0.6561** |
| | RMSE | 5.2590 | 5.6460 | 4.7490 | 4.7750 | 4.5710 | **1.1064** |
| Two | MAPE | 0.2833 | 0.2085 | 0.1771 | 0.1728 | 0.1638 | **0.0545** |
| | MAE | 6.8030 | 6.4010 | 5.6550 | 5.6030 | 5.4230 | **1.4588** |
| | RMSE | 10.1470 | 10.0590 | 9.1990 | 9.3440 | 8.9890 | **2.0545** |
| Three | MAPE | 0.4447 | 0.4186 | 0.2713 | 0.2714 | 0.2421 | **0.0665** |
| | MAE | 10.0460 | 10.1000 | 8.3700 | 8.4380 | 8.0900 | **1.7446** |
| | RMSE | 14.6170 | 14.4840 | 13.6720 | 13.6490 | 13.2420 | **2.4648** |

### 4.3 Statistic test

Notably, relying on the above evaluation criteria, i.e., MAPE, MAE and RMSE, alone for determining the best forecasting model is not statistically efficient. As such, we go a step further and test all our out-of-sample forecasting results for statistical significance using both the modified Diebold-Mariano (DM) test.

In particular, the DM test statistically confirms the superiority of CEEMDAN-DeepTCN over benchmarks, under a confidence level of 99% (see **Table 5**). Two interesting findings can be found. First, when the proposed CEEMDAN-DeepTCN model is treated as the testing target, the *p*-values are all far smaller than 1%, indicating that the proposed deep learning forecasting model performs statistically better than all other models in all cases, under the confidence level of 99%. Second, according to **Table 5**, all the three deep learning models (i.e., DeepTCN, GRU and LSTM) can yield better results than the other two forecasting models at the significance level of 1%, verifying the superiority of the deep learning to the time series model and artificial neural network. Third, the BPNN can achieve a better performance than the LR model, under the confidence level of 95%.

**Table 5.** Results of the DM test between proposed approach and benchmarks (*p*-value).

|      | CEEMDAN-DeepTCN | DeepTCN | GRU | LSTM | BPNN |
|------|-----------------|---------|-----|------|------|
| DeepTCN | -42.72 (0.00) | | | | |
| GRU | -43.86 (0.00) | -4.36 (0.00) | | | |
| LSTM | -47.14 (0.00) | -4.52 (0.00) | -0.19 (0.85) | | |
| BPNN | -43.19 (0.00) | -19.61 (0.00) | -17.49 (0.00) | -17.51 (0.00) | |
| LR | -45.40 (0.00) | -22.00 (0.00) | -20.45 (0.00) | -19.58 (0.00) | -2.37 (0.02) |

**4.4 Robustness analysis**

In this section, the standard deviation of the all three evaluation criteria of CEEMDAN-DeepTCN (i.e., MAPE, MAE and RMSE) and the comparison among CEEMDAN with the benchmark prediction models are conducted to prove the robustness of the proposed hybrid forecasting framework. First, in terms of the standard deviation, we ran the proposed model CEEMDAN-DeepTCN ten times, and calculated the corresponding standard deviations, and the detail of robustness analysis with standard deviation is demonstrated in **Table 6**. Second, in terms of decomposition-based method, the proposed hybrid forecasting framework still makes a best performance at all horizons (see **Table 7**).

The above results are demonstrated in **Tables 6-7**, and support evidence that (1) the robustness of the proposed hybrid prediction results is verified; (2) the deep learning methods provide a better forecasting performance than the time series model and the artificial intelligence model.

**Table 6.** Robustness analysis with standard deviation

| Horizon | Criteria | Mean (Std) |
|---------|----------|------------|
| One | MAPE | 0.026 (0.001) |
|  | MAE | 0.653 (0.043) |
|  | RMSE | 1.029 (0.062) |
| Two | MAPE | 0.054 (0.001) |
|  | MAE | 1.362 (0.045) |
|  | RMSE | 1.929 (0.067) |
| Three | MAPE | 0.068 (0.001) |
|  | MAE | 1.788 (0.043) |
|  | RMSE | 2.556 (0.111) |

Table 7. Performance comparison of different methods with CEEMDAN in terms of criteria

| h | Criteria | Forecasting models | | | | |
|---|---|---|---|---|---|---|
| | | CEEMDAN-LR | CEEMDAN-BPNN | CEEMDAN-LSTM | CEEMDAN-GRU | CEEMDAN-DeepTCN |
| One | MAPE | 0.0468 | 0.0387 | 0.0311 | 0.0305 | **0.0265** |
| | MAE | 1.0575 | 0.9933 | 0.7812 | 0.7991 | **0.6561** |
| | RMSE | 1.4950 | 1.4971 | 1.2892 | 1.3521 | **1.1064** |
| Two | MAPE | 0.0927 | 0.0665 | 0.0642 | 0.0623 | **0.0545** |
| | MAE | 1.9898 | 1.7935 | 1.6721 | 1.6253 | **1.4588** |
| | RMSE | 2.6645 | 2.5782 | 2.4363 | 2.3635 | **2.0545** |
| Three | MAPE | 0.1610 | 0.0883 | 0.0839 | 0.0810 | **0.0665** |
| | MAE | 3.1754 | 2.3280 | 2.1671 | 2.0534 | **1.7446** |
| | RMSE | 4.0633 | 3.2105 | 3.0572 | 2.8863 | **2.4648** |

## 4.5 Summary

From the empirical study, we can obtain the following four major conclusions.

(1) The proposed hybrid forecasting framework, i.e., CEEMDAN-DeepTCN, significantly outperform all the considered benchmarking methods in terms of all the three evaluation criteria for the $PM_{2.5}$.

(2) The powerful predictive power of weather variables and time variables can be verified statistically, by employing the embedding technology map the categorical variables of time and weather into vectors in $PM_{2.5}$ prediction.

(3) By constructing the deep learning technique (i.e., DeepTCN), the discrete variables' pattern can be effective modelled and can be used as an efficient tool for analyzing and forecasting the complex system, such as oil price.

(4) The results also reveal the superiority of CEEMDAN-Deep over the prevailing deep learning techniques, artificial neural network and time series method, in capturing the related variables' data pattern, thereby guaranteeing the prediction accuracy.

## 5. Conclusions

Due to the intrinsic complexity of $PM_{2.5}$ data in terms of its interactive involving factors, a novel hybrid forecasting method (i.e., CEEMDAN-DeepTCN), considering the related pollutant concentration variables, the meteorological factors and time variables, is proposed for $PM_{2.5}$ forecasting. The empirical study shows that the employed decomposition technique and the deep learning forecasting method can significantly improve prediction performance and statistically outperform some other popular forecasting methods (including the traditional time series model, artificial neural network, and the popular deep learning models) in terms of prediction accuracy and robustness. This further indicates that the proposed hybrid forecasting method (i.e., CEEMDAN-DeepTCN), which can effective capture and model the data patterns of the $PM_{2.5}$ and the related exogenous variables, can be used as a very promising method for complex time series forecasting problems, especially for $PM_{2.5}$ with high volatility, irregularity and more interactive involving factors.

Besides $PM_{2.5}$, the proposed forecasting method (i.e., CEEMDAN-DeepTCN) can be also used for other tough forecasting tasks. Furthermore, this study only considers univariate time series analysis without processing of time series. Therefore, some decomposition methods (such as SSA, EEMD and VMD methods) can be also taken into consideration to enhance the prediction capability of the novel model, by extracting the different timescale of the $PM_{2.5}$. We will look into these issues in the near future.

## Conflicts of interest

The authors declare that there are no conflicts of interest regarding the publication of this study.

## Acknowledgments

This research work was partly supported by the National Natural Science Foundation of China under Grant No. 71988101 and by the Fundamental Research Funds for the Central Universities under Grant No. xpt012020022. This research work was also partly supported by the Academic Excellence Foundation of BUAA for PhD Students, and by the Innovation Ability Improvement Project in Colleges and Universities of Gansu Province of China under Grant No. 2019A-060.